\documentclass[preprintnumbers,showpacs,twocolumn,amsmath,amssymb]{revtex4}

\usepackage{amsmath}
\usepackage{amsfonts}
\usepackage{amssymb}
\usepackage{verbatim}
\usepackage{graphicx}
\usepackage[caption=false]{subfig}
\usepackage{color}

\newcommand{\beq}{\begin{equation}}
\newcommand{\eeq}{\end{equation}}
\newcommand{\bqa}{\begin{eqnarray}}
\newcommand{\eqa}{\end{eqnarray}}

\newcommand{\smallfrac}[2]{\mbox{$\frac{#1}{#2}$}}

\newcommand{\ket}[1]{ |{#1} \rangle}

\definecolor{ngreen}{rgb}{0.2,0.6,0.2}

\definecolor{golden}{rgb}{0.8,0.6,0.1}

\definecolor{purp}{rgb}{0.8,0.1,0.8}

\definecolor{orange}{rgb}{0.9,0.3,0}
\definecolor{mar}{rgb}{0.6,0.1,0.1}

\definecolor{oxblue}{RGB}{0., 33, 71}
\definecolor{camblue}{RGB}{163,193,173}

\begin{document}

\title{A Cavity-Enhanced Room-Temperature Broadband Raman Memory}

\author{D. J. Saunders$^{1,\dag}$, J. H. D. Munns$^{1,2}$, T. F. M. Champion$^1$, C. Qiu$^{1,3}$,  K.~T.~Kaczmarek$^1$, E. Poem$^{1}$, P. M. Ledingham$^1$, I. A. Walmsley$^1$, J. Nunn$^1$}

\affiliation{$^{1}$Clarendon Laboratory, University of Oxford, Parks Road, Oxford OX1 3PU, United Kingdom\\
$^{2}$ QOLS, Blackett Laboratory, Imperial College London, London SW7 2BW, UK\\
$^{3}$Department of Physics, Quantum Institute for Light and Atoms, State Key Laboratory of Precision Spectroscopy, East China Normal University, Shanghai 200062, People's Republic of China
}

\date{\today}

\begin{abstract} 
Broadband quantum memories hold great promise as multiplexing elements in future photonic quantum information protocols. Alkali vapour Raman memories combine high-bandwidth storage, on-demand read-out, and operation at room temperature without collisional fluorescence noise. However, previous implementations have required large control pulse energies and suffered from four-wave mixing noise. Here we present a Raman memory where the storage interaction is enhanced by a low-finesse birefringent cavity tuned into simultaneous resonance with the signal and control fields, dramatically reducing the energy required to drive the memory. By engineering anti-resonance for the anti-Stokes field, we also suppress the four-wave mixing noise and report the lowest unconditional noise floor yet achieved in a Raman-type warm vapour memory, $(15\pm2)\times10^{-3}$ photons per pulse, with a total efficiency of $(9.5\pm0.5)$\%. 

\end{abstract}


\maketitle

Quantum information technologies such as quantum key distribution and random number generators are beginning to transition into the commercial sphere, where key requirements are the ability to function at high speed, and in non-laboratory settings. The high carrier frequency of optical signals enables photonic quantum devices to operate noise-free at room temperature whilst offering GHz-THz operational bandwidths. However, direct photon-photon interactions are prohibitively weak, which has held back the development of photonic quantum processors. One solution to this problem has been to use probabilistic measurement-induced non-linearities \cite{OBrien:2009}, but the probability of success decreases exponentially with system size, limiting the state of the art to $<10$ photons \cite{Yao:2012}. Further scaling photonic devices requires a multiplexing strategy. Quantum memories capable of storing photons and releasing them on-demand provide the ability to temporally multiplex a repeat-until-success architecture to achieve a freely scalable photonic quantum information platform operable at room temperature.

There are many types of quantum memory for light under development: electromagnetically induced transparency \cite{Chen:2013aa}, the full atomic-frequency comb protocol \cite{Mustafa:2015,Pierre:2015}, gradient-echo memories \cite{Hedges:2010,Hosseini:2011}, and the far-off-resonant Raman memory \cite{Reim:2011aa,England:2013aa}. Each of these protocols have advantages and challenges, see Ref.~\cite{MemoryReview:2010} for a recent review. A helpful metric for temporal multiplexing is the time-bandwidth product $B=\tau \delta$, with $\delta$ the acceptance bandwidth and $\tau$ the memory lifetime. $B$ is the maximum number of time-bins over which a memory can synchronise an input signal. Time-bandwidth products of $B>1000$ enable a dramatic enhancement in the multiphoton rate from parametric photon sources \cite{Nunn:2013}, as required to utilise multiphoton interference for computational speed-up (e.g. boson sampling \cite{Aaronson:2011}). Large time-bandwidth products have been achieved with cold atom memories \cite{Bao:2012aa,Dudin:2013aa}, cryogenic rare earth ion memories \cite{Heinze:2013} and room-temperature Raman memories \cite{Reim:2011aa,Bustard:2013aa}. In this paper we present an implementation of an alkali-vapour Raman memory, operating at $75^\circ$C, and show the lowest-noise floor demonstrated in such a room-temperature memory to date.

In a Raman memory, an intense control field induces a two-photon absorption feature far-detuned from any atomic resonance. The absorption linewidth of the field-dressed atoms can be made broadband with a pulsed control \cite{Bustard:2013aa}, enabling a large time-bandwidth product even with $\sim\mu$s coherence times in a room-temperature vapour \cite{Reim:2011aa}, and allowing direct interfacing with pulsed heralded photon sources \cite{Michelberger:2015,England:2015aa}. Operation with temporally short and far-detuned pulses also removes contamination of the retrieved signals by collision-induced fluorescence \cite{Manz:2007aa}. However, large control pulse energies are needed to drive the storage interaction far from resonance, and the control can also drive unwanted four-wave mixing, introducing noise which cannot be filtered out either spectrally or temporally. Four-wave mixing noise has emerged as the last remaining roadblock to the development of efficient $\Lambda$-type room-temperature quantum memories \cite{Phillips:2008aa,Lauk:2013aa,Vurgaftman:2013,Dabrowski:2014aa,Michelberger:2015}. We solve this problem by introducing a new \emph{cavity enhanced} Raman memory protocol. We place the atoms inside a low-finesse optical cavity that both enhances the strength of the Raman interaction, reducing the power requirements on the control field, and also suppresses four-wave mixing. 

\begin{figure*}[t!]
\includegraphics[width=1\linewidth]{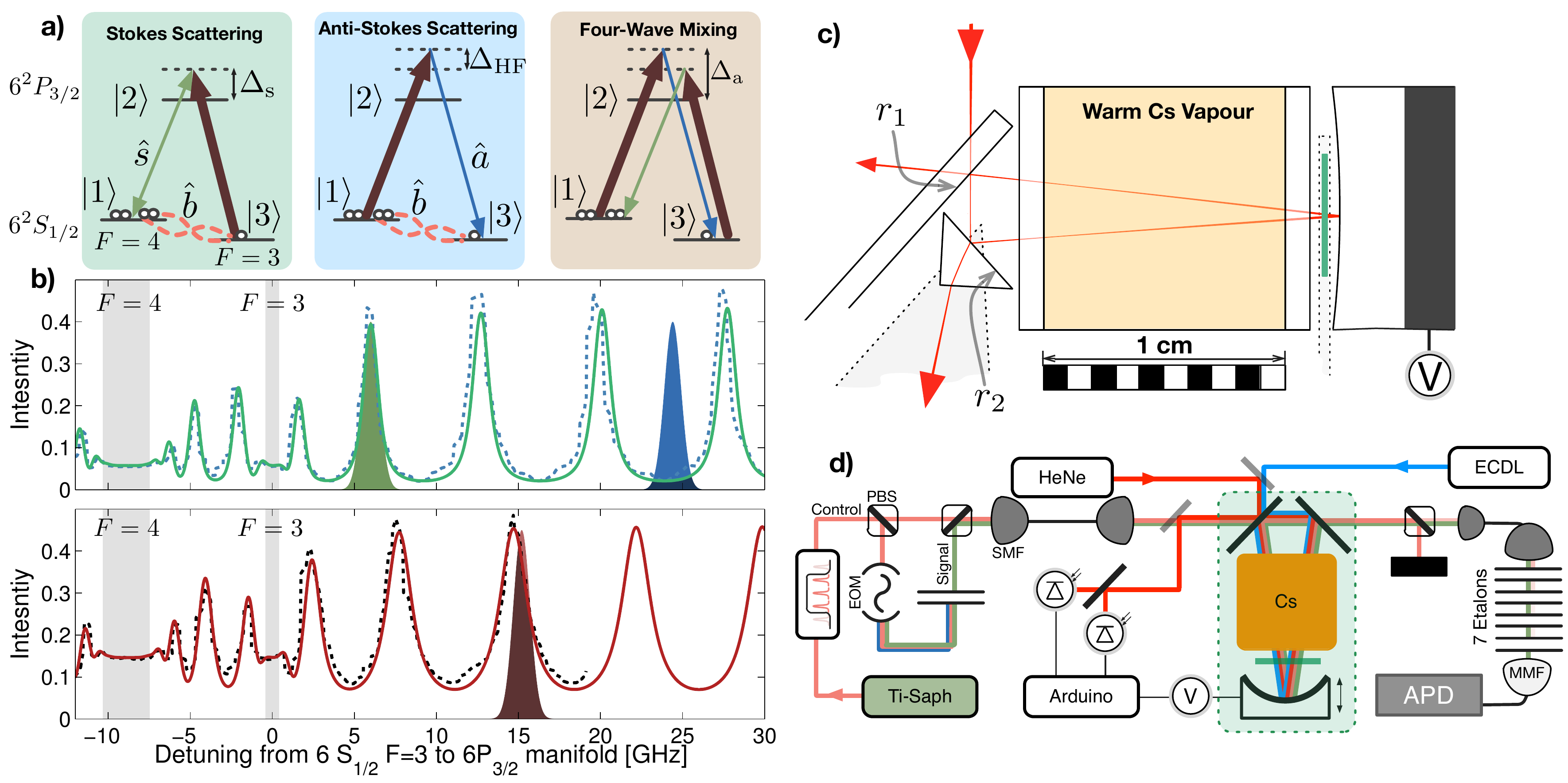}

\caption{(a) Energy level diagrams showing the D2 $\Lambda$ level scheme in Cs. There are two major interactions: one, the Stokes scattering --- involving the signal field $\hat s$ (green arrow), control field (red arrow), and spin-wave $\hat b$ (red loop), this is the desired memory interaction; two, anti-stokes scattering --- involving the control field, anti-stokes mode $\hat a$ (blue arrow), and spin-wave $\hat b$; the third diagram is the effective noise-process, anti-stokes scattering followed by stokes-scattering, a four-wave-mixing interaction. (b) Measured frequency response of the atom-filled cavity system. The top is for the signal and anti-stokes polarisation, and the bottom is for the control polarisation. The dashed lines are measured responses (see supplementary material) and the solid lines are our theory model using measured values of the cavity parameters~\cite{Munns:2015}. The grey solid areas represent the absorption in the cavity, and are an indication of our spin-polarisation, with $\sim80\%$  in the ideal $F=4$ state. The green pulse shows the location of the signal, the blue the anti-stokes, and the red the control. (c) Scale diagram of the cavity Raman quantum memory. We operate the ring-cavity in transmission, where in-coupling and out-coupling mirrors have amplitude reflectivities $r_1$ and $r_2$ respectively. The curved mirror has $R=40$mm, and the small green rectangle represents the $200\mu$m-thick LiNbO$_3$ birefringent crystal which we tune with temperature to reach the desired cavity conditions. (d) Experimental schematic, for more details see the main text and the supplementary material.}
\label{fig:Cavity}
\end{figure*}

To describe how the cavity enhances the memory, we first introduce the standard Raman protocol, which we have developed in caesium vapour at $\sim70^\circ$C \cite{Reim:2010aa,Reim:2011aa,England:2012aa,Sprague:2014aa,Michelberger:2015}, using the $\Lambda$-system of the Cs D2 line comprising the hyperfine ground states $6 S_{1/2}$, $F=3$ ($\ket{3}$), $F=4$ ($\ket{1}$), with a hyper-fine splitting of $\Delta_\mathrm{HF}=9.2$~GHz and the $6P_{3/2}$ excited state manifold ($\ket{2}$), whose hyperfine structure is unresolved, see Fig.~\ref{fig:Cavity} (a). We initialise the memory by optically pumping the atoms into $\ket{1}$. Light storage and subsequent retrieval are mediated by the application of bright control pulses with bandwidth $\delta$, time-dependent Rabi frequency $\Omega(t)$ and detuning $\Delta_\mathrm{s}$ from resonance with the $\ket{3}\leftrightarrow\ket{2}$ transition. This adiabatically couples the signal mode $\hat s$ to a delocalised excitation of the storage state $\ket{3}$ described by the spin-wave mode $\hat b$, according to the beam-splitter-like Hamiltonian $\mathcal{H}_\mathrm{BS}\propto C_\mathrm{s}\hat s \hat b^\dagger + \mathrm{h.c.}$, where the Stokes coupling parameter is given by $C_\mathrm{s} = \sqrt{Wd\gamma}/\Delta_\mathrm{s}$, with $W = \int |\Omega(t)|^2\mathrm{d}t$ proportional to the control pulse energy $\mathcal{E}$ and $d$ the resonant optical depth of the $\ket{1}\leftrightarrow\ket{2}$ transition, with linewidth $\gamma$ \cite{Nunn:2007aa}. The origin of the unwanted four-wave mixing noise is the coupling of the control field to the ground-excited transition $\ket{1}\leftrightarrow\ket{2}$ with detuning $\Delta_\mathrm{a}=\Delta_\mathrm{s}+\Delta_\mathrm{HF}$, on which it drives spontaneous anti-Stokes scattering in mode $\hat a$, producing anti-Stokes photons and spin-wave excitations in pairs \cite{Wasilewski:2006aa}. The interaction is accordingly described by a two-mode-squeezing Hamiltonian $\mathcal{H}_\mathrm{TMS}\propto  C_\mathrm{a} \hat a \hat b + \mathrm{h.c.}$, with the anti-Stokes coupling given by $C_\mathrm{a} = C_\mathrm{s}\Delta_\mathrm{s}/\Delta_\mathrm{a}$. If $\Delta_\mathrm{s} \gtrsim \Delta_\mathrm{HF}$, the interactions described by $\mathcal{H}_\mathrm{BS}$, $\mathcal{H}_\mathrm{TMS}$ are of comparable strength, and the probability of spontaneously generating a spin-wave excitation approaches that of mapping an incident signal photon into the spin-wave. The retrieved fields are then contaminated with thermal noise (half a two-mode-squeezed state), which destroys the non-classical statistics of stored single photons \cite{Michelberger:2015}.

One way to prevent four-wave mixing is to use polarisation selection rules such that the control field only couples to the $\ket{3}-\ket{2}$ transition. However, for alkali atoms this is not possible due to destructive interference between interaction pathways associated with different intermediate excited states $\ket{2}$ \cite{Vurgaftman:2013}. Four-wave mixing can also be suppressed in dispersive media where it is poorly phasematched \cite{England:2013aa,England:2015aa}, but the Stokes shift $\Delta_\mathrm{HF}$ in alkali vapours is not large enough to introduce a significant phase mismatch. A third approach towards suppressing four-wave mixing is the creation of a second Raman absorption feature at the anti-stokes frequency \cite{Romanov:2015}. Whilst promising, this method requires the use of an atomic species with three ground states, and therefore cannot be implemented in $^{133}$Cs.

The cavity-enhanced Raman memory instead suppresses $\mathcal{H}_\mathrm{TMS}$ relative to $\mathcal{H}_\mathrm{BS}$ by engineering the density of scattering states so that anti-Stokes scattering into the fundamental cavity mode is forbidden by destructive interference. That is, we arrange for the signal (Stokes) field to coincide with a cavity resonance, while the anti-Stokes field is anti-resonant with the cavity. Setting the free-spectral range of the cavity accordingly, with $\mathrm{FSR}_m = 4\Delta_\mathrm{HF}/(2m+1)$, for integer $m$, and a roundtrip optical path length $l_m = c/\mathrm{FSR}_m$, the spontaneous generation of spurious spin wave excitations by anti-Stokes scattering via $\mathcal{H}_\mathrm{TMS}$ is then reduced relative to $\mathcal{H}_\mathrm{BS}$ by the ratio of intra-cavity field amplitudes, parameterised by the noise suppression factor $x = (1-\mu_\mathrm{s})/(1+\mu_\mathrm{a})\approx 0.24$, where $\mu_{\mathrm{s,a},\Omega}$ denotes the cavity-roundtrip amplitude losses for the intra-cavity fields for the Stokes ($\mu_\mathrm{s} = 0.6$), anti-Stokes ($\mu_\mathrm{a} = 0.6$) and control ($\mu_\Omega = 0.4$) frequencies. Much greater noise suppression could be achieved with a moderate increase in the cavity quality factor.

With a cavity around the atoms, the acceptance bandwidth of the memory is now limited by the cavity resonance linewidth, so that a broadband memory requires a low-finesse cavity. However, unlike in single-atom cavity QED \cite{Specht:2011aa}, where very high cavity quality factors are required to achieve the strong coupling regime and suppress spontaneous emission into any direction, here we require only the suppression of on-axis spontaneous anti-Stokes scattering, since off-axis scattering couples to momentum-orthogonal spin wave modes that are not phasematched \cite{Dabrowski:2014aa} and do not contribute noise. Cavity-enhanced ensemble Raman storage is therefore technically less demanding than cavity-QED-based storage, requiring only low- to moderate-finesse, which remains compatible with broadband operation ($\delta\sim$ GHz).

The broadest acceptance bandwidth is obtained for a zero-order cavity with $\mathrm{FSR}_0=36.8$~GHz, but to achieve $l_0 = 8.1$~mm would require a monolithic construction. For our demonstration, we instead constructed a second-order cavity with $\mathrm{FSR}_2=7.36$~GHz and $l_2=40.8$~mm. We chose a ring geometry instead of a linear one so as not to produce a standing wave inside the cavity, since atoms diffusing into the field nodes during the storage time would not interact at retrieval. To in-couple the control field, which is linearly polarised orthogonal to the signal and anti-Stokes modes, we tune the cavity birefringence by heating a thin sliver of LiNbO$_3$ (see Fig.~\ref{fig:Cavity}) to achieve simultaneous resonance for the signal and the control fields. The cavity length is stabilised with a H\"{a}nsch-Couillaud lock \cite{Hansch:1980}: we inject a HeNe laser into the cavity and feed back on the polarisation rotation of the beam by adjusting the cavity length with a piezo-electric actuator (Fig.~\ref{fig:Cavity}). An Arduino microcontroller (Due) provides both fast (kHz) and slow (mHz) feedback (see Supplementary Material).

Besides suppressing FWM noise, the double resonance configuration for the control and signal fields reduces the control pulse energy required to drive the memory by a factor $\sim \mathcal{F}^2_\mathrm{s}\mathcal{F}^2_\Omega/\pi^4$ compared to an equivalent free-space memory, where $\mathcal{F}_{\mathrm{s},\Omega}$ denotes the cavity finesse for the signal and control modes. In our experiment with $\mathcal{F}_\mathrm{s}\sim 7$ and $\mathcal{F}_\Omega\sim 4$. This combined with sleigh imperfect resonance of the control field (Fig.~\ref{fig:Cavity} (b)), we obtained a reduction in energy by $\sim 10$; a moderate increase in cavity quality could achieve extremely low operating power.

\begin{figure}[h!]
\begin{center}
\includegraphics[width=.9\linewidth]{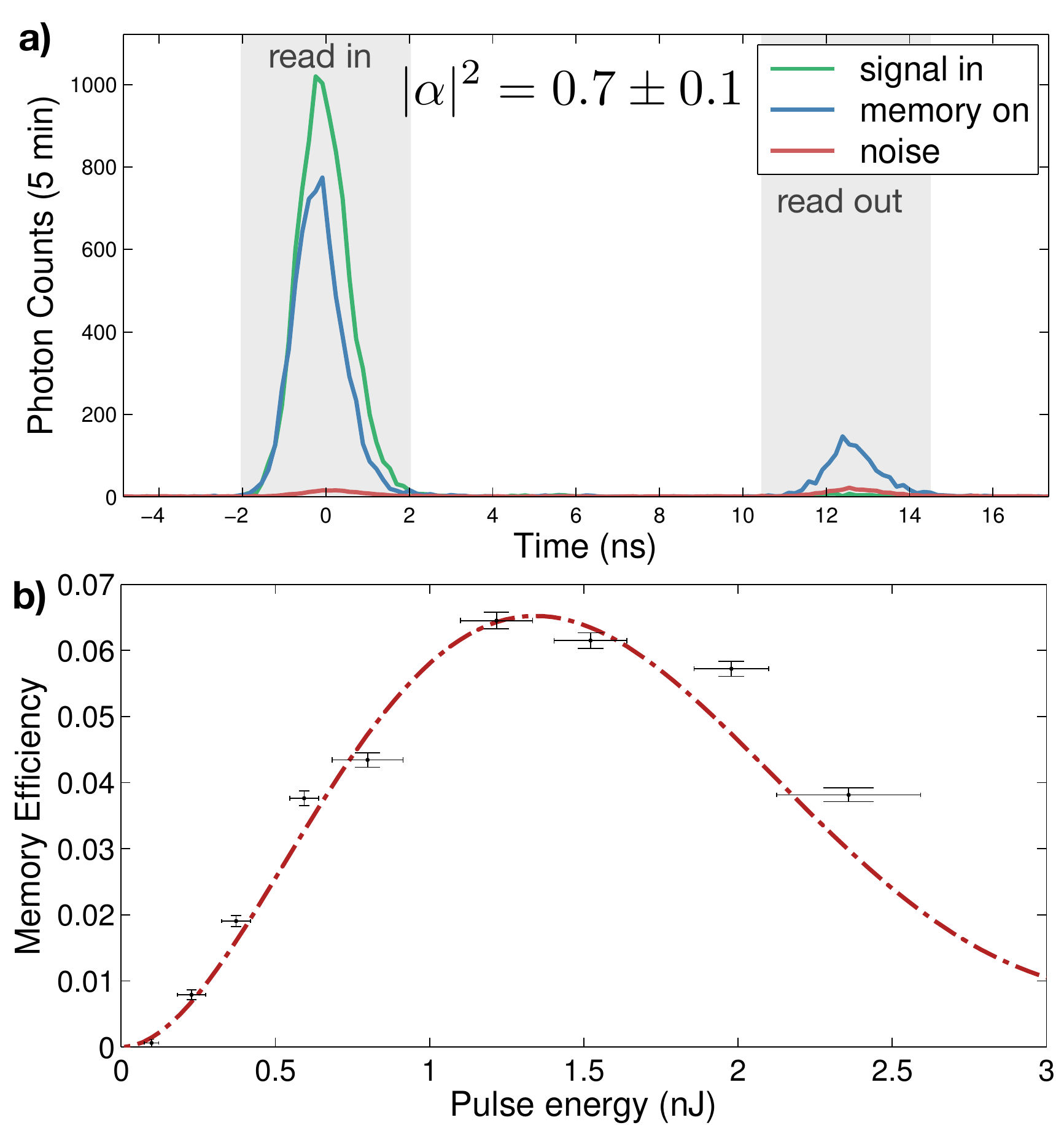}
\end{center}
\caption{(a) Arrival time histograms for the signal field emerging from the cavity with the control pulses blocked (\emph{signal in}), unblocked (\emph{memory on}) and with the input signal field blocked (\emph{noise}). This is measured for a coherent state amplitude of $|\alpha^2|=0.7\pm0.1$. (b) Measured (points) and predicted (line) memory efficiency, as a function of the intra-cavity control pulse energy.}
\label{fig:effplot} 
\end{figure}

As can be seen from Fig.~\ref{fig:Cavity}(b), absorption and dispersion around the atomic resonances significantly modify the cavity mode structure \cite{Munns:2015}. The Cs vapour pressure depends exponentially on temperature, so to maintain the resonance condition required for the memory we developed active temperature stabilisation to within $0.1^\circ$C, see supplementary material for details.

The experimental setup for the memory is shown in Fig.~\ref{fig:Cavity}(d). The signal and control fields are derived from a modelocked Ti:Sapph oscillator producing pulses with $320$~ps duration ($\delta=1.2$GHz); details can be found in \cite{Michelberger:2015}.
The signal and control pulses are coupled into the same single-mode fibre and directed towards the cavity memory. After the memory we block the control pulses with polarisation and spectral filtering, with a rejection of $120$~dB and a transmission for the signal of $~1\%$. We detect the transmitted and retrieved signal photons with a Geiger-mode avalanche photodiode (Perkin-Elmer SPCM-AQRH) and time-to-digital convertor with $81$ps resolution (qutools quTAU). Fig~\ref{fig:effplot}(a) shows storage and retrieval with consecutive control pulses separated by $12.5$~ns. The input signal average photon number is $0.7\pm0.1$ and the memory efficiency is $(9.5\pm0.5)\%$, while the noise floor is $\langle n^\mathrm{noise} \rangle = 0.015\pm0.002$ photons per control pulse in the read-out time bin, the lowest noise floor measured in warm vapour Raman memories to date. We also measure the effect of memory efficiency with control pulse energy, Fig~\ref{fig:effplot}(b). We observe a maximum efficiency at $\mathcal{E}\sim1.5$nJ, before the memory efficiency is decreased due to the dynamical AC-stark effect. Nevertheless, the measured efficiency agrees with our new theoretical model, see supplementary material and ref. \cite{josh_theory}. This effect can be compensated by using appropriate pulse shaping \cite{Gorshkov:2007aa,Phillips:2008aa,Reim:2010aa}. We also measured the memory lifetime to be $\tau=95\pm7$ns, without any magnetic shielding (see supplementary information). This agrees with our previous measurements without shielding and we expect to easily extend this to $>\mu s$ with magnetic field shielding in the future \cite{Reim:2011aa}. 

\begin{figure}[t!]
\begin{center}
\includegraphics[width=.9\linewidth]{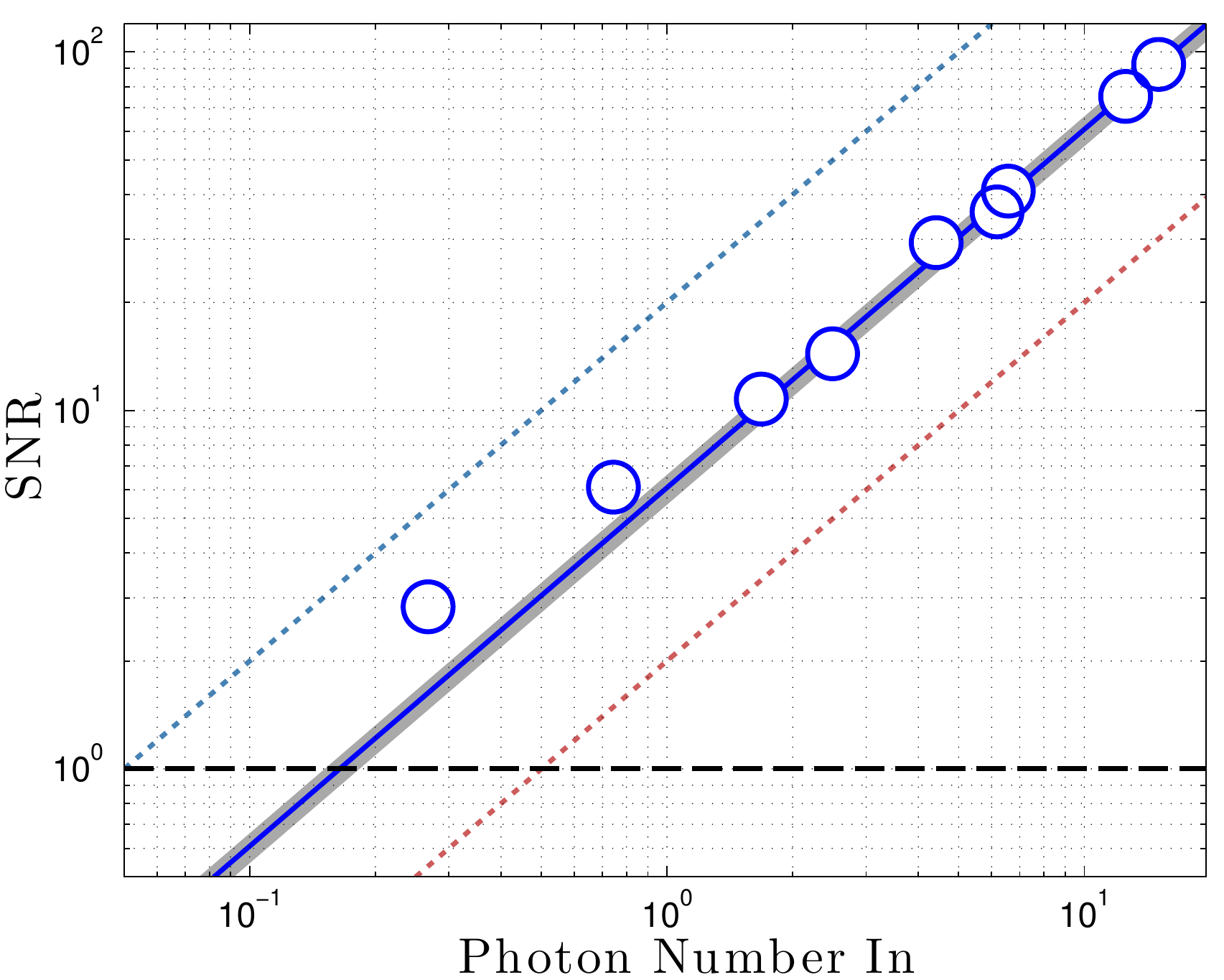}
\end{center}
\caption{Measurement of $\mu_1$. To measure $\mu_1$ we operate the memory with a control pulse energy of $\sim 1.4$~nJ, corresponding to our optimal SNR condition. We then measure the signal-to-noise ratio for various input coherent state amplitudes. Where the fits cross the horizontal-black dashed line (SNR=1) is the value of $\mu_1$; with smaller $\mu_1$ desirable. The blue circles are our measured values; error bars are smaller than marker dimension. The lower-red dashed line represents the state-of-the-art using cavity-free Raman memory experiments \cite{Michelberger:2015}, with a $\mu_1'\approx0.5$. The upper-blue dashed line is the expected performance from our theoretical model, with $\mu_1^\mathrm{Ideal}=0.005$. We do not reach this ideal noise-floor due to experimental imperfections not included the model (see supplementary material). After fitting our data with $y=x/\mu_1$, we extract $\mu_1=0.17\pm0.02$. The error is calculated using a Monte-Carlo simulation.}
\label{fig:noise}
\end{figure}

The noise floor in our experiment contains contributions from four-wave mixing and also from spontaneous emission. To compare with the theory, we fit our measurements and extract the FWM $\mathcal{E}^2$-scaling component of the noise floor to find $0.006 \pm 0.003$, which agrees well with the theoretical prediction of $0.005$ (Supplementary Information). Furthermore, a convenient metric to determine if we have suppressed the noise is the $\mu_1$ parameter \cite{Mustafa:2015}, where $\mu_1=\langle n^\mathrm{noise} \rangle / \eta^\mathrm{tot}$; a fair metric to compare different devices because it includes both memory efficiency and noise floor. To determine this accurately, we vary the coherent state amplitude and determine the SNR for each, shown in Fig 3. By fitting this data, we determine a raw $\mu_1$ of $0.17\pm0.02$, a considerable improvement compared to free-space Raman memories, for which $\mu_1'\sim0.5$ \cite{Michelberger:2015}. This is a clear demonstration of noise suppression in a far-off resonant broadband Raman memory for the first time.

In order to take full advantage of the cavity-induced FWM noise suppression, it will be necessary to remove the remaining sources of noise. This can be achieved by improving the optical pumping efficiency, and suppression of counts arising from the pumping laser. The task at hand therefore reduces to a readily solvable engineering challenge. Once achieved, we can predict the expected performance of a future cavity Raman memory with heralded single photon inputs. Even at the current moderate memory efficiencies, and using SPDC sources which have previously been shown to be compatible with broadband Raman memories \cite{Michelberger:2015}, interfacing them with our new experiment should enable non-classical revival of single-photons \cite{josh_theory}. A realistic route to increasing the memory efficiency beyond $10\%$ is to compensate for the AC-stark shift using pulse shaping, increasing the caesium number density, or increasing the finesse. Therefore, our new cavity enhanced Raman memory has laid the ground work for a true room-temperature quantum memory for broadband heralded single photons in the near future.

We thank Andreas Eckstein for assistance with the time-tagger; and Benjamin Brecht, Amir Feizpour and Steve Kolthammer for insightful discussions. This work was supported by the UK Engineering and Physical Sciences Research Council (EPSRC; EP/J000051/1 and Programme Grant EP/K034480/1), the Quantum Interfaces, Sensors, and Communication based on Entanglement Integrating Project (EU IP Q-ESSENCE; 248095), the Air Force Office of Scientific Research: European Office of Aerospace Research \& Development (AFOSR EOARD; FA8655-09-1-3020), EU IP SIQS (600645), the Royal Society (to J.N.), and EU Marie-Curie Fellowships PIIF-GA-2013-629229 to D.J.S. and PIEF-GA-2013-627372 to E.P. I.A.W. acknowledges an ERC Advanced Grant (MOQUACINO). C. Qiu is supported by the China Scholarship Council.

\clearpage 
\onecolumngrid

\setcounter{figure}{0}
\renewcommand{\thefigure}{S\arabic{figure}}

\begin{center}
\large{\bf Supplementary Material}
\end{center}

\section*{Experimental set-up}

The control and signal pulses are both derived from a mode-locked Tsunami titanium-sapphire (Ti:Saph) laser, producing an 80-MHz train of pulses with a centre wavelength of 852 nm and FWHM bandwidth of 1.2 GHz. The Ti:Sa frequency is actively stabilised at a detuning of $\Delta=15.2$ GHz relative to the $6^{2}S_{1/2}(F=3)$ $\leftrightarrow$ $6^{2}P_{3/2}$ transition of a Cs reference cell. A Pockels cell (Quantum Technology QC) is used to select pulses for the storage and retrieval processes, setting an experimental repetition rate of $8$ kHz. The signal field is generated with an electro-optic modulator (EOM) which is designed to shift the frequency of the modulated sidebands by an amount equal to the Cs ground-state hyperfine splitting $\Delta_{\mathrm{HF}}$. The red-detuned sideband, which satisfies the two-photon resonance condition with the control field, is frequency-selected using an air-spaced Fabry-Perot etalon. The orthogonally polarised signal and control pulses are temporally and spatially overlapped and coupled into a two-junction ring cavity in a co-propagating configuration. In addition, a narrowband Toptica DL pro external cavity diode laser (ECDL) is introduced into the cavity in a counter-propagating fashion in order to prepare the atoms in the initial ground state $\ket{1}$ via optical pumping.

\begin{figure}[ht!]
\begin{center}
\includegraphics[width=.9\linewidth]{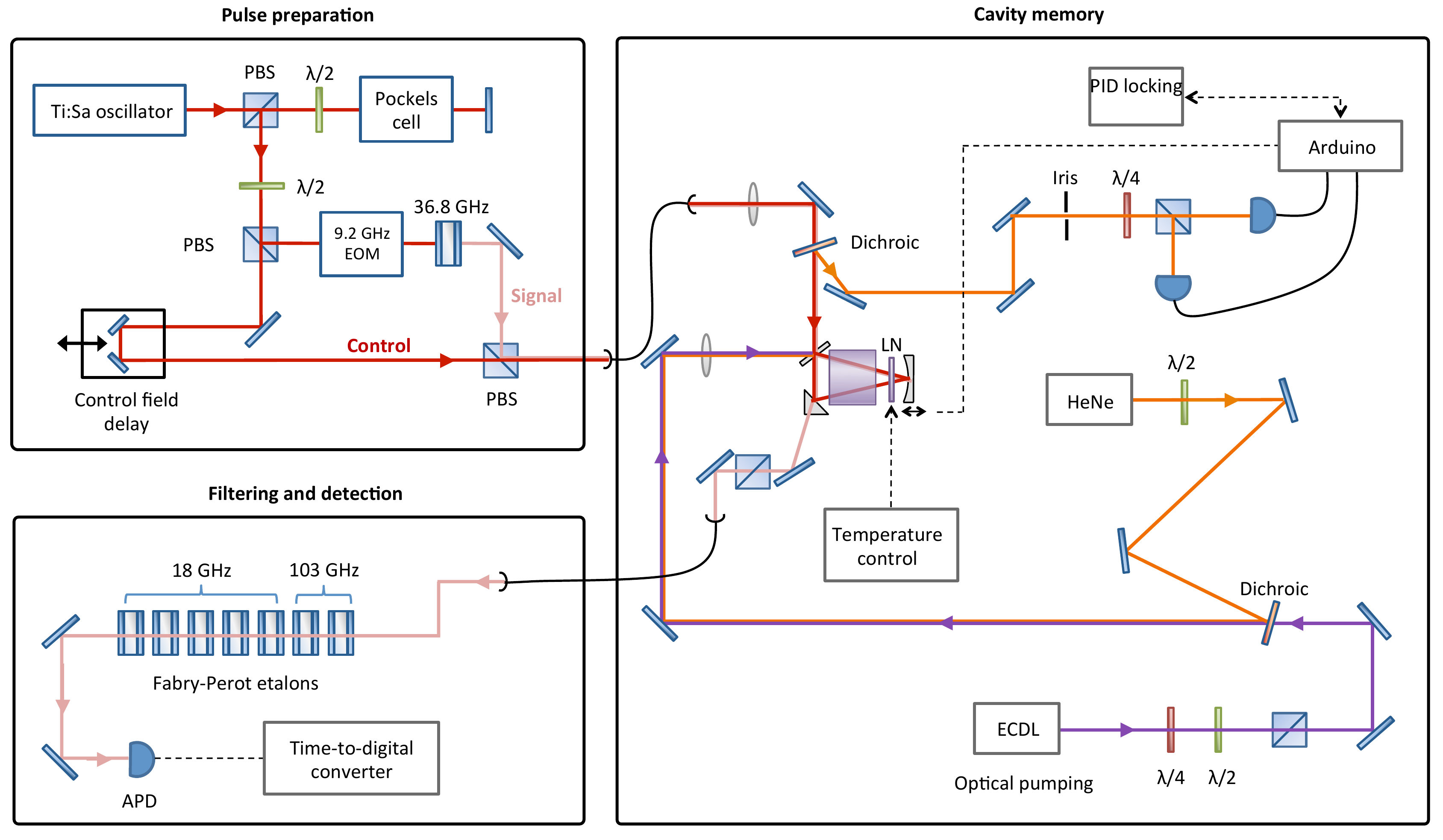}
\end{center}
\caption{Detailed schematic of the experiment. The experiment is separated into three parts: one, the pulse preparation; two, the cavity memory; and three, the filtering and detection. To prepare the pulses we pick pulses from our Ti:Sa oscillator and then split them on a PBS. From this PBS the two output paths are matched such that when they recombine on the second PBS they are overlapping temporally. We use an EOM to create the signal pulses, see main text for details. After this the signal and control pulses are coupled into the same fibre and sent to the cavity memory. On the cavity memory table we align all four fields, the pumping laser, the control and signal, and the locking laser to our ring-cavity. The output of the cavity is first filtered using a Glan-Taylor PBS before being coupled into single-mode-fibre and being sent to the filtering stage. Here the signal is passed through an effective total of 7 Fabry-Perot Etalons, with a control extinction of $>90$dB before being coupled in multi-mode-fiber and coated on an APD and our time-tagging system. The total transmission of the measurement stage is $\sim 2$\%.}
\label{fig:setup}
\end{figure}

The cavity length is actively stabilised using a variant of the H\"{a}nsch-Couillaud method whereby an error signal is generated by monitoring the difference signal between orthogonal polarisation components of the HeNe locking laser transmitted through the cavity, arising from the relative phase accumulated on passing through the birefringent crystal, monitored by the 12-bit ADC of a microcontroller (Arduino Due) which then provides feedback to  a piezo-electric ring actuator (Noliac) which controls the position of the concave cavity mirror along the axis of symmetry.   As a result of the periodicity of the error signal with cavity length being significantly different to the cavity response at the signal wavelength, it was necessary for the cavity length to remain locked to the correct set point solution of the error signal (corresponding to maximal signal coupling):  in order to ensure this was the case, the feedback to the actuator consisted of two components: (1) a fast PID calculation, based on the Arduino library \cite{PID}, with an output spanning the voltage range corresponding to a length change of the cavity associated with one period at the signal wavelength, and  (2) a slow offset component which increments when the PID output approaches 90\% of its range, spanning the whole voltage range of the actuator.  Due to the finite precision of the DAC of the microcontroller, the two output components were output seperately and combined on a voltage adding circuit in the appropriate ratio to maximise the sensitivity of the output. This set up allows the lock to follow slower period drifts of the cavity whilst still remaining stable on the kHz timescale, and enabled the cavity to remain locked to within \mbox{$\pm50\,\mathrm{Hz}$} for a duration of hours.

The temperature of the Cs vapour cell is \mbox{$84.3^\circ\mathrm{C}$}. This ensemble has both a large measured pressure broadening (\mbox{$\mathrm{FWHM} = 80\,\mathrm{MHz}$}), from 10~Torr of Ne buffer gas, and a measured Doppler broadening (\mbox{$\mathrm{FWHM=375}\,\mathrm{MHz}$}). This makes both numerical modelling challenging \cite{Munns:2015}. This puts tight bounds on temperature stability, with changes of less than a degree causing large GHz size shifts in the cavity response; easily moving us away from memory operation. To solve this we developed a temperature control system to keep the temperature to within $0.1$ of the target temperature. We monitored the temperature of the cell using thermistors and another Arduino Due, then using LabView we implemented a PID loop controlling the voltage to the vapour cells heating wires. The last step is to ensure the cavity is well aligned and has good fringe visibility --- the key parameter in FWM suppression. To measure this we scan the cavity length using the piezo mirror, enabling us to measure $I_{\text{max}}(f)$ and $I_{\text{min}}(f)$; the maximum and minimum transmission intensities of the cavity for a given frequency, $f$. The cavity response in these measurements is affected by the linewidth of our probe laser ($\delta = 1.2$GHz). We measure visibilities, $V=\smallfrac{I_{\text{max}}-I_{\text{min}}}{I_{\text{max}}+I_{\text{min}}}$, of $(86\pm2)\%$ for the Stokes frequency and polarisation, and $(71\pm5)\%$ for the Control frequency and polarisation, and $(86\pm5)\%\%$ for the anti-Stokes frequency and polarisation. The differences between these visibilities are ascribed to the change in mirror reflectivity for the control and signal, anti-Stokes polarisations, and a small amount of residual linear absorption at the signal frequency \cite{Munns:2015}.

To ensure we are operating the memory in the optimal \emph{triple resonance} condition --- Stokes resonant, control resonant, anti-Stokes anti-resonant --- we lock the cavity to the Stokes frequency and then scan our Ti-Saph laser from the D2-line to 30GHz blue detuned; limited by the $\sim30$GHz mode-hop-free tuning range. We then measure the control polarisation, and tune the temperature of the birefringent element to make the control field also resonant. An example cavity frequency scan is shown in Figure 1b. Reaching triple resonance was made challenging because of dispersion induced by the Cs ensemble, as such we developed a full theoretical model, see \cite{Munns:2015}.

The signal field transmitted through the cavity is coupled into a single-mode optical fibre and isolated from the control field using a series of polarisation and frequency filters. The arrival time of detection events relative to a trigger signal are recorded using a time-to-digital converter (qutools quTAU) with a time bin resolution of 81 ps.

\section*{Memory efficiency and memory lifetime}

\begin{figure}[t!]
\begin{center}
\includegraphics[width=.5\linewidth]{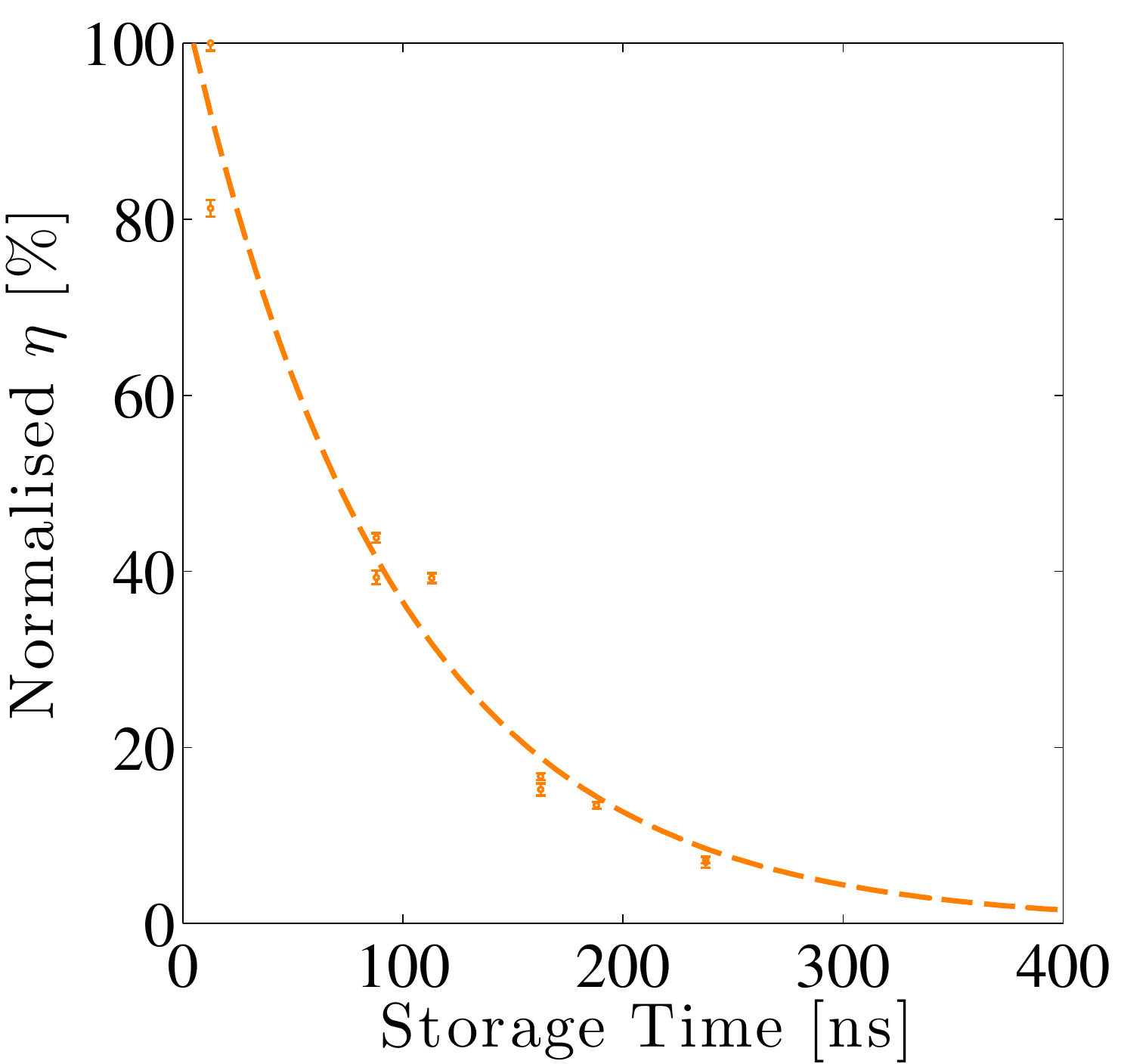}

\end{center}
\caption{We measured the memory lifetime by storing input pulses with average photon number $\sim 10$ and recording the average number of retrieved photons, for a range of storage times. We extract a memory lift time of $95\pm7$~ns from the exponential decay.}
\label{fig:lifetime}
\end{figure}

The memory efficiency can be evaluated from the rates of detection events $c_{l}^\mathrm{in/out}$ in the read-in and read-out time bins with $l \in \{sc, s, c\}$ denoting the combination of signal and control fields present. Note that the optical pumping laser, which is used to initiate the atomic ensemble prior to each storage event, is periodically switched off during the memory interaction to avoid depletion of the stored spin-wave excitation. The total memory efficiency for storage followed by retrieval is given by 

\begin{equation}\label{eq:cav_eff}
\eta_{\mathrm{tot}} = \frac{c_{\mathrm{sc}}^{\mathrm{out}} - c_{\mathrm{c}}^{\mathrm{out}}- c_{\mathrm{s}}^{\mathrm{out}}}{c_{\mathrm{s}}^{\mathrm{in}}},
\end{equation}

\noindent Equation \eqref{eq:cav_eff} assumes that the number of counts in the signal mode can be determined by subtracting the noise counts $c_{cd}^\mathrm{in/out}$ in each time bin. It is worth clarifying this assumption: The measured detection events consist of contributions from the signal (Stokes) mode, the anti-Stokes mode and the control mode. The latter contribution is the result of residual leakage of the control field through the filtering stage. We have implicitly assumed that the amount of control leakage is independent of the presence of the signal field, i.e. these leakage counts cancel in the expression $c_{\mathrm{scd}}^{\mathrm{out}} - c_{\mathrm{cd}}^{\mathrm{out}}$. The count rates $c_{l}^\mathrm{in/out}$ are found by integrating the pulse traces in Figure \ref{fig:effplot} (a) within the range indicated by the shaded areas. This gives a total memory efficiency of $\eta_{\mathrm{tot}} = 9.5 \pm 0.5\%$. 

To measure the memory lifetime we vary the time between our read-in and read-out control pulses using our pulse picker, see \cite{Reim:2011aa} for details. The results are shown in Fig. \ref{fig:lifetime}.

\section*{Theory of cavity Raman Interaction}

The theory of cavity-enhanced $\Lambda$-type quantum memories has been treated in detail \cite{Dantan:2005aa,Gorshkov:2007aa}. We model the efficiency and noise in our Raman cavity memory by extending these analyses to include both Stokes and anti-Stokes fields \cite{josh_theory}. Fig.~\ref{fig:effplot}(c) shows good agreement between the measured memory efficiencies and the prediction of our theoretical model, with no free parameters, given by 
$$
\eta_\mathrm{tot} = |\chi C_\mathrm{s} E \kappa |^2,
$$
where $\chi = (1-r^2)/(1-\mu_\mathrm{s})$ is the amplitude transmission of the signal field with input- and output-coupling mirror amplitude reflectivities $r_1 = r_2 =  r\approx 0.86$. $\kappa = (e^\zeta E)^{-1/2}(1-e^{-f})/f$ is the input signal overlap with the cavity response, where $\zeta = -2\Re\{f\}$ and $E = (1-e^{-\zeta})/\zeta$, with
$$
f = -C_\mathrm{s}^2 -C_\mathrm{a}^2 x+\mathrm{i} \frac{W}{\Delta_\mathrm{s}} + \mathrm{i} \frac{W}{\Delta_\mathrm{a}}.
$$
Since the intra-cavity fields pass through the atoms many times, the optical depth $d$ is replaced by the cavity cooperativity $\mathcal{C} = rd/(1-\mu_\mathrm{s})\approx \mathcal{F}d/\pi$, where in our experiment the cavity finesse is $\mathcal{F}\approx 7$. Operationally, a measurement of the visibility $V=90\%$ of the cavity fringes in Fig.~\ref{fig:Cavity} provides an estimate of the noise suppression via the relation $x^2 = (1-V)/(1+V)$. Our model then predicts the noise floor of the memory to be
$$
N_\mathrm{noise} = |\chi C_\mathrm{s}C_\mathrm{a}x|^2\times \frac{1-e^{-\zeta}E}{\zeta}
$$
photons per pulse. Using a measured beam-waist of $w=150~\mu$m, we calculate $W[\textrm{GHz}]\approx 110\times \mathcal{E}[\textrm{nJ}]$. We measure $V = 90\%$ and infer a resonant single-round-trip optical depth $d = 300$ with a pressure-broadened linewidth $\gamma = 250$~MHz \cite{Munns:2015}. 

We then obtain the prediction $N_\mathrm{noise} = 0.005$ photons per pulse for the maximum efficiency shown in Fig.~\ref{fig:effplot}(a).

\section*{Calculation of FWM Contribution}

\begin{figure}[h!]
\begin{center}
\includegraphics[width=.5\linewidth]{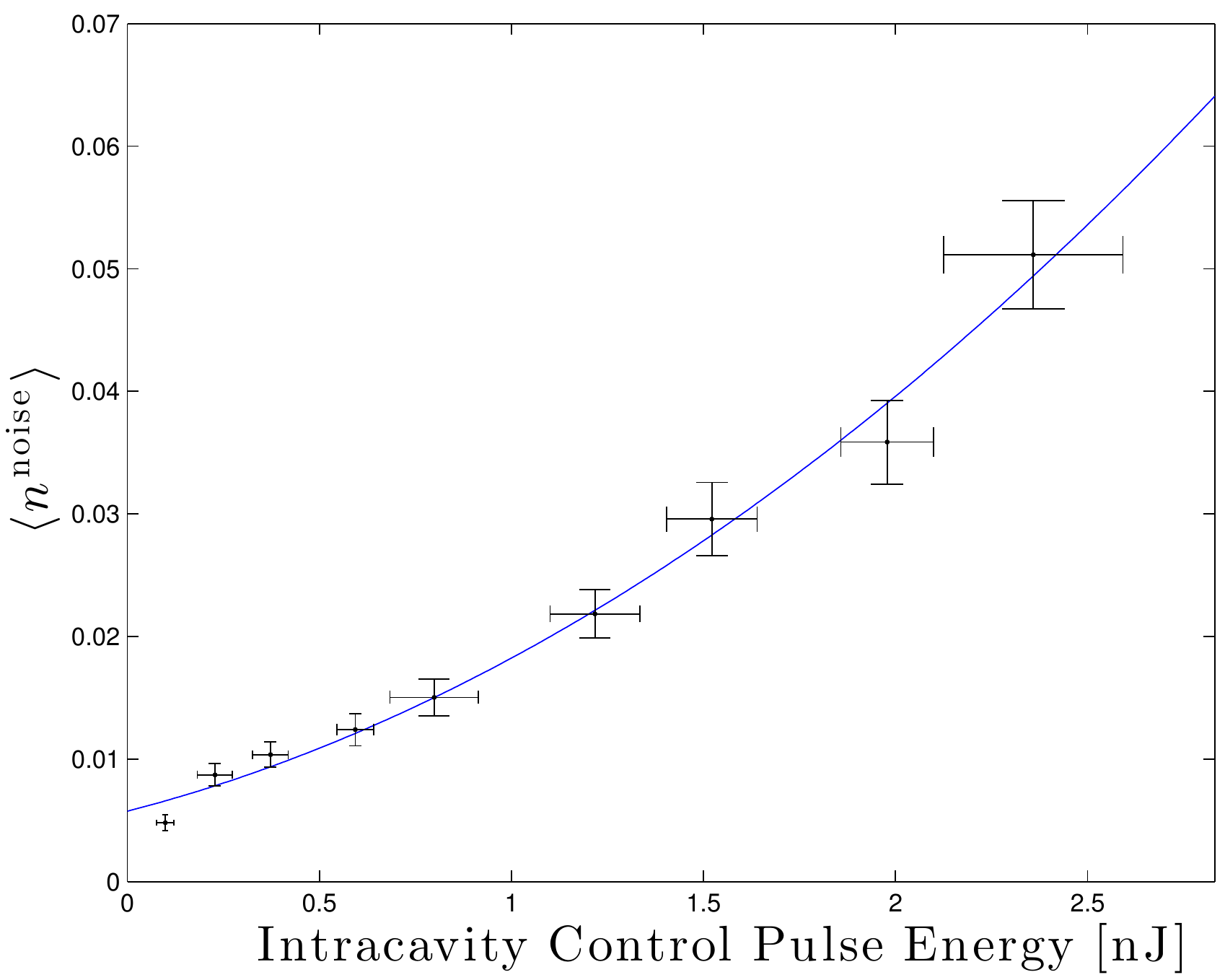}
\end{center}
\caption{Measured scaling of the noise-floor in our cavity memory. }
\label{fig:noise_scaling}
\end{figure}

In order to isolate the contribution to the noise floor from FWM, we measured the scaling of the noise-floor as a function of the control pulse energy $\mathcal{E}$. The FWM contribution is expected to scale quadratically with $\mathcal{E}$, since two control pulse photons need to scatter to produce each FWM noise photon. The fit reveals the presence of a constant noise background and a component linear in $\mathcal{E}$. These two terms arise from control-field-independent noise, such as dark counts or fluorescence, and spontaneous Raman scattering from un-pumped population in the storage state, respectively. By extracting the quadratic term, we find a FWM contribution to the measured noise floor of $0.006\pm0.003$ for the control-field energy corresponding to our optimal SNR, which agrees with the prediction of our model (see above). With technical improvements to our optical pumping, our detection stage and our buffer gas (to further reduce collisional noise), we would be limited only by the FWM noise, which as we have shown can be removed with an appropriate cavity design.

\end{document}